\documentclass[twocolumn,prl,aps,superbib,tightenlines,floatfix]{revtex4}
\usepackage{amsfonts}
\usepackage{amsmath}
\usepackage{amssymb}
\usepackage{graphicx}
\begin{document}
\topmargin -24mm
\textheight 244mm
\bibliographystyle{apsrev}
\title{Theory of local heating in nanoscale conductors}
\author{Yu-Chang Chen}
\author{Michael Zwolak}
\author{Massimiliano Di Ventra~\cite{MD}}
\affiliation{Department of Physics, Virginia Polytechnic Institute and State University,
Blacksburg, Virginia 24061}
\begin{abstract}
We report first-principles calculations of local heating in nanoscale
junctions formed by a single molecule and a gold point contact. Due to the lower 
current density and larger heat 
dissipation, the single molecule heats up less than the gold
point contact. We also find that, at zero temperature, a threshold bias $%
V_{onset} $ of about 6 mV and 11 mV for the molecule and the point contact,
respectively, is required to excite the smallest vibrational mode and
generate heat. The latter estimate is in very good agreement with recent
experimental results on the same system. At a given external bias $V$ below $%
V_{onset}$, heating becomes noticeable when the background temperature is on
the order of $\sim e(V_{onset}-V)/k_{B}$. Above $V_{onset}$, local heating
increases dramatically with increasing bias but is also considerably
suppressed by thermal dissipation into the electrodes. The results provide a
microscopic picture of current-induced heat generation in atomic-scale
structures.
\end{abstract}
\pacs{73.63.Nm, 68.37.Ef, 73.40.Jn}
\maketitle
Local heating occurs when electrons diffusing in a conductor release energy
to the ions via scattering with phonons. The amount of heat generated in a
given portion of the conductor depends on several factors: the strength of
the electron-phonon interaction, the current density, the background
temperature, and the electron mean free path. In a nanoscale junction, the
electron mean free path is large compared to the dimensions of the junction.
As a consequence, each electron releases only a small fraction of its energy
during the time it spends in the junction.~\cite{todorov1,todorov2,todorov3}
However, there can still be substantial local heating due to the large
current density in nanoscale junctions. Indeed, such an effect has recently
been observed in several atomic-scale structures.~\cite{vieira,muller,van}
Apart from its fundamental importance in solid state physics, the problem
has thus gained renewed interest due to its possible impact on nanoscale
electronics. At this length scale, all atoms of the junction and
corresponding vibrational modes need to be treated explicitely, and the
electronic distribution calculated self-consistently with the correct
scattering boundary conditions. Only a few theoretical investigations,
mainly using the tight-binding approximation, have tackled this problem at
the atomic level.~\cite{todorov1,todorov2,todorov3}

In this letter, we derive a general expression for the power transferred by
the electrons to the vibrational modes of a nanoscale junction and calculate
the local heating using first-principles approaches. Combined with a model
of heat transfer to the electrodes, the approach provides a microscopic and
detailed picture of current-induced heat generation in nanoscale conductors.
As an example, we discuss local heating for a single molecule between gold
electrodes and a gold point contact. We find that the single-molecule
junction heats up less than the gold point contact due to a larger heat
dissipation into the electrodes and larger resistance to electron flow. In
the case of the gold point contact, experimental data is available which is
in very good agreement with the predicted threshold bias, $V_{onset}$, for
heat generation and corresponding local temperature. At a given external
bias $V$ below $V_{onset}$, we find that local heating becomes noticeable
when the background temperature is larger than $\sim e(V_{onset}-V)/k_{B}$,
but is also considerably suppressed by thermal dissipation into the
electrodes.

Let us start by considering a nanoscale structure between two bulk
electrodes. The many-body Hamiltonian of the system is 
\begin{equation}
H=H_{el}+H_{ion}+H_{el-ion},  \label{hamiltonian}
\end{equation}
where $H_{el}$ is the electronic Hamiltonian, $H_{ion}=\sum_{i=1}^{N}\mathbf{%
P}_{i}^{2}/2M_{i}+\sum_{i<j}V_{ion}(\mathbf{R}_{i}- \mathbf{R}_{j})$ is the
ionic Hamiltonian, and $H_{el-ion}=\sum _{i,j}V_{el-ion}(\mathbf{r}_{i}-%
\mathbf{R}_{j})$ describes the electron-ion interaction. $\mathbf{R}_{i}$, $%
\mathbf{P}_{i}$, and $M_{i}$ are the coordinates, momentum, and mass,
respectively, of the i-th ion. $\mathbf{r}_{i}$ is the coordinate of the
i-th electron. In the adiabatic approximation, the Hamiltonian $\left( \ref%
{hamiltonian}\right)$ has effective single-particle eigenvalues $%
\Psi_{E}^{L(R)}\left( \mathbf{r,\mathbf{K}_{\Vert}}\right)$ with energy E
and momentum $\mathbf{K}_{\Vert}$ parallel to the electrode surface,
corresponding to electrons incident from the left (right) electrodes.~\cite%
{diventra1} We calculate the single-particle wavefunctions using a
scattering approach within the density functional theory of many-electron
systems.~\cite{diventra1}

In order to calculate the electron-phonon interaction, we consider a small
deviation $\mathbf{Q}_{i}=\mathbf{R}_{i}-\mathbf{R}_{i}^{0}$ of the i-th ion
from its equilibrium position $\mathbf{R}_{i}^{0}$. We introduce normal
coordinates $\left\{ q_{i\mu }\right\} $ such that the $\mu $-th component ($%
\mu =x,y,z$) of $\mathbf{Q}_{i}$ is 
\begin{equation}
\left( \mathbf{Q}_{i}\right) _{\mu }=\sum_{j=1}^{N}\sum_{\nu =1}^{3}A_{i\mu
,j\nu }q_{j\nu }.  \label{normal}
\end{equation}%
The transformation matrix, $\mathbf{A}=\left\{ A_{i\mu ,j\nu }\right\} $,
satisfies the orthonormality relations: $\sum_{i,\mu }M_{i}A_{i\mu ,j\nu
}A_{i\mu ,j^{\prime }\nu ^{\prime }}=\delta _{j\nu ,j^{\prime }\nu ^{\prime
}}$. The Hamiltonian describing the ionic vibrations can then be decoupled
into a set of independent harmonic oscillators:

\begin{equation}
H_{vib}=\frac{1}{2}\sum\limits_{i,\mu \in vib.}\dot{q}_{i\mu }^{2}+\frac{1}{2%
}\sum\limits_{i,\mu \in vib.}\omega _{i\mu }^{2}q_{i\mu }^{2},  \label{osc}
\end{equation}%
where $\left\{ \omega _{i\mu }\right\} $ are the normal mode frequencies and
the summations are carried out for all normal modes. We calculate these
modes from first-principles.~\cite{sayvetz}

\begin{figure}[tbp]
\includegraphics[width=.48\textwidth]{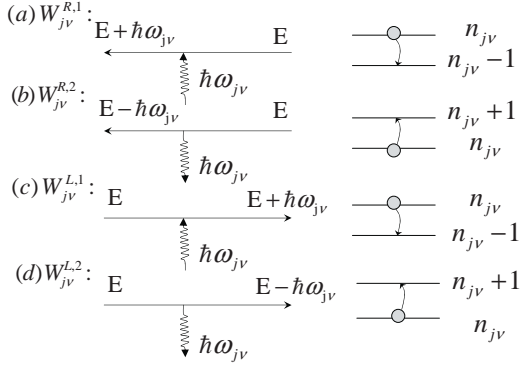}
\caption{Feynman diagrams of the four main electron-phonon scattering
mechanisms contributing to local heating of the junction. (a) Cooling
process due to absorption of a phonon from a left-moving electron. (b)
Heating process due to the emission of a phonon from a left-moving electron.
(c) and (d) are the equivalent mechanisms corresponding to the right-moving
electrons.}
\label{feynman}
\end{figure}

We now introduce the field operator, $\hat{\Psi}=\hat{\Psi}^{L}+\hat{\Psi}%
^{R}$, which describes electrons incident from the left (L) and right (R)
electrodes, where~\cite{chen} 
\begin{equation}
\hat{\Psi}^{L(R)}=\sum_{E} a_{E}^{L(R)} \Psi_{E}^{L(R)}\left( \mathbf{r,%
\mathbf{K}_{\Vert}}\right) .  \label{field}
\end{equation}
The coefficients $a_{E}^{L(R)}$ are the annihilation operators for electrons
incident from the left (right) electrode. They satisfy the usual
anti-commutation relations $\{a_{E}^{i},a_{E^{\prime}}^{j\dagger}\}=\delta
_{ij}\delta\left( E-E^{\prime}\right) $. We also assume that the electrons
rapidly thermalize into the bulk electrodes so that their statistics are
given by the equilibrium Fermi-Dirac distribution, $f_{E}^{L(R)}=1/(\exp[%
(E-E_{FL(R)})/k_{B}T_{e}]+1)$, in the left (right) electrodes, where $T_{e}$
is the background temperature.

We can now express $\left( \mathbf{Q}_{i}\right) _{\mu }$ in terms of boson
annihilation and creation operators. Together with the field operator $\hat{%
\Psi}$, the second-quantized form of $H_{vib}$ and the electron-vibration
interaction, $H_{el-vib}$, can be written as 
\begin{equation}
H_{vib}=\sum_{j\nu \in vib.}(b_{j\nu }^{\dag }b_{j\nu }+\frac{1}{2})\hbar
\omega _{j\nu },  \label{shm}
\end{equation}%
and 
\begin{align}
H_{el-vib}& =\sum_{\alpha ,\beta }\sum_{E_{1},E_{2}}\sum_{i\mu ,j\nu \in
vib.}  \notag \\
& \sqrt{\frac{\hbar }{2\omega _{j\nu }}}A_{i\mu ,j\nu }J_{E_{1},E_{2}}^{i\mu
,\alpha \beta }a_{E_{1}}^{\alpha \dag }a_{E_{2}}^{\beta }\left( b_{j\nu
}+b_{j\nu }^{\dag }\right) ,  \label{hei}
\end{align}%
where $\alpha ,\beta =L,R$ ; $i,j=1,...,N$; and $\mu ,\nu =1,2,3$. The
quantity $J_{E_{1},E_{2}}^{i\mu ,\alpha \beta }$ is the electron-vibration
coupling constant and has the form 
\begin{equation}
J_{E_{1},E_{2}}^{i\mu ,\alpha \beta }=\int d\mathbf{r}\int d\mathbf{K}%
_{\Vert }\Psi _{E_{1}}^{\alpha \ast }\left( \mathbf{r},\mathbf{K}_{\Vert
}\right) \partial _{\mu }V^{ps}\left( \mathbf{r},\mathbf{R}_{i}\right) \Psi
_{E_{2}}^{\beta }\left( \mathbf{r},\mathbf{K}_{\Vert }\right) ,
\label{couplej}
\end{equation}%
where we have chosen to describe the electron-ion interaction with the
pseudopotentials $V^{ps}\left( \mathbf{r},\mathbf{R}_{i}\right) $ for each
i-th ion.~\cite{diventra1} $b_{j\nu }$ are the annihilation operators for
the $j\nu $-th vibrational modes and satisfy the commutation relation $\left[
b_{j\nu },b_{j^{\prime }\nu ^{\prime }}^{\dag }\right] =\delta _{j\nu
,j^{\prime }\nu ^{\prime }}$. The statistics of these modes are described by
the Bose-Einstein distribution: 
\begin{equation}
\left\langle n_{j\nu }\right\rangle =1/\left[ \exp \left( \hslash \omega
_{j\nu }/k_{B}T_{w}\right) -1\right] ,  \label{bose}
\end{equation}%
where $T_{w}$ is the local temperature of the junction.

For each normal mode, we can now calculate the rates of thermal energy
generated by the electron-vibration interactions corresponding to the
first-order processes described in Fig.~\ref{feynman}. These processes
correspond to electrons incident from the right or left electrode that
absorb (cooling) or emit (heating) energy because of the electron-vibration
scattering. We evaluate these rates with the Fermi golden rule 
\begin{align}
W_{j\nu }^{R,k}& =2\pi \hbar \left( \delta _{k,2}+\left\langle n_{j\nu
}\right\rangle \right) \int dE\left\vert \sum_{i\mu }A_{i\mu ,j\nu }J_{E\pm
\hbar \omega _{j\nu },E}^{i\mu ,LR}\right\vert ^{2}  \notag \\
\cdot & f_{E}^{R}(1-f_{E\pm \hbar \omega _{j\nu }}^{L})D_{E\pm \hbar \omega
_{j\nu }}^{L}D_{E}^{R},  \label{pmode1}
\end{align}

\begin{equation}
W_{j\nu}^{L,k}=W_{j\nu}^{R,k}(R\rightleftharpoons L),  \label{pmode2}
\end{equation}
where $"R\rightleftharpoons L"$ means interchange of labels R and L; the
positive (negative) sign in Eq. $\left( \ref{pmode1}\right)$ is for $k=1(2)$%
, corresponding to relaxation (excitation) of vibrational modes; $%
D_{E}^{L(R)}$ is the partial density of states corresponding to $%
\Psi_{E}^{L(R)}$, whose sum is the total density of states. The term $\delta
_{k,2}$ corresponds to spontaneous emission. Finally, a factor of 2 due to
spin degeneracy appears in Eq. $\left( \ref{pmode1}\right)$.

Since electrons can excite all possible energy levels of a mode with
frequency $\omega _{j\nu }$, the statistical average $\left\langle n_{j\nu
}\right\rangle $ is required. The total thermal power generated in the
junction is therefore the sum over all vibrational modes for the four
processes of Fig. $\left( \ref{feynman}\right) $: 
\begin{equation}
P=\sum_{j\nu \in vib.}\left( W_{j\nu }^{R,2}+W_{j\nu }^{L,2}-W_{j\nu
}^{R,1}-W_{j\nu }^{L,1}\right)   \label{power}
\end{equation}

\begin{figure}[tbp]
\includegraphics[width=.48\textwidth]{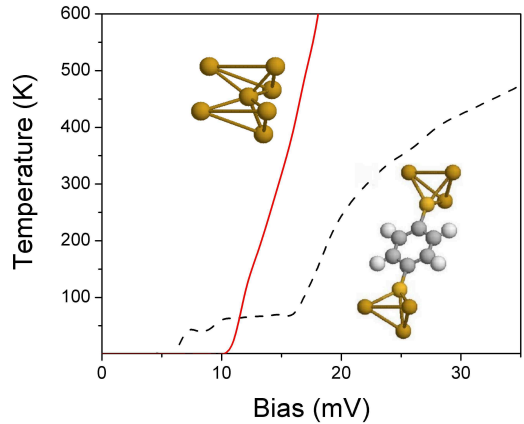}
\caption{Local equilibrium temperature as a function of bias for a
benzene-dithiolate molecular junction (dashed line, schematic in the
right-lower corner) and a single-gold-atom point contact (solid line,
schematic in the left-upper corner) at $T_{e}=0$ K. No heat dissipation into
the electrodes is taken into account.}
\label{eqtemp}
\end{figure}

Equation $\left( \ref{power}\right)$ is the central result of this paper. It
allows for a first-principles calculation of the local temperature in a
nanoscale junction. We now discuss results for two specific cases: local
heating in a benzene-dithiolate molecular junction and a gold point contact
(see schematics in Fig.~\ref{eqtemp}). We also assume the left electrode to
be positively biased. At zero temperature (i.e., $T_{e}=0$ and $T_{w}=0$),
the heating processes $\left\{W_{j\nu}^{L,2}\right\}$, corresponding to
electrons incident from the left electrode, vanish due to the Pauli
exclusion principle; the cooling processes $\left\{ W_{j\nu}^{R,1}\right\} $%
\ and $\left\{ W_{j\nu}^{L,1}\right\}$, corresponding to the transitions
from a high to a low energy level of the modes, are also prohibited because
all modes are at the ground state. The only nonzero contribution to local
heating is therefore from $W_{j\nu }^{R,2}$, i.e.

\begin{align}
P& =2\pi \hbar \sum_{j\nu \in vib.}\left( 1+\left\langle n_{j\nu
}\right\rangle \right)  \notag \\
& \cdot \int_{E_{FL}+\hbar \omega _{j\nu }}^{E_{FR}}dE\left\vert \sum_{i\mu
}A_{i\mu ,j\nu }J_{E-\hbar \omega _{j\nu },E}^{i\mu ,LR}\right\vert
^{2}D_{E-\hbar \omega _{j\nu }}^{L}D_{E}^{R}.  \label{zero}
\end{align}

A bias greater than $V_{onset}=\min\{\hbar\omega_{j\nu}\}/e$ is therefore
necessary to generate heat. The smallest vibrational frequency calculated
from first principles~\cite{sayvetz} is about 6 mV and 11 mV for the
molecule and the point contact, respectively. The latter estimate is in very
good agreement with recent experimental observations.~\cite{vieira} When the
bias is smaller than the threshold voltage, local heating is only possible
at nonzero background temperature $T_{e}$. This is caused by a small
fraction of thermally excited electrons which can induce level transitions
of the normal modes. The heating generated is substantial when $%
k_{B}T_{e}\approx\min\{\hbar\omega_{j\nu}\}-(E_{FR}-E_{FL})$.

At $T_{e}=0$ and $T_{w}> 0$, the junction heats up progressively, and
eventually an equilibrium temperature is reached when the heating processes (%
$W_{j\nu }^{R,2}$\ and $W_{j\nu}^{L,2}$) balance the cooling processes ($%
W_{j\nu}^{R,1}\ $and $W_{j\nu}^{L,1}$). 
The corresponding local equilibrium temperature is plotted in Fig.~\ref%
{eqtemp} for both the single molecule junction and the gold point contact.
It is evident from Fig.~\ref{eqtemp} that the equilibrium temperature
increases abruptly above the threshold voltage and it is already substantial
at biases of only few mV. The reason for this dramatic increase resides
again in the Pauli exclusion principle which suppresses the cooling
processes considerably.

\begin{figure}[tbp]
\includegraphics[width=.48\textwidth]{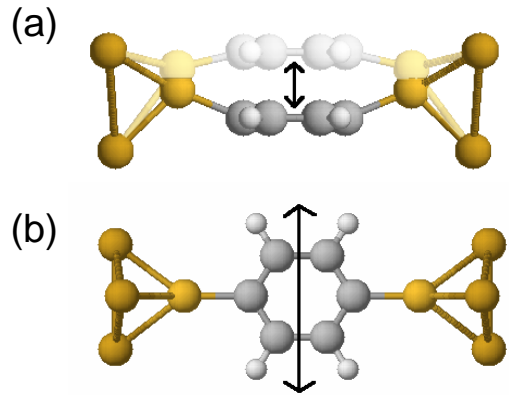}
\caption{The two modes contributing to heating at $T_{e}=0$ and $V<10$ mV.
(a) The vibrational mode with energy $\hbar \protect\omega \approx 6$ meV.
(b) The mode with energy $\hbar \protect\omega \approx 9$ meV.}
\label{modes}
\end{figure}

In the case of the gold point contact, the longitudinal vibrational mode of
the single gold atom ($\hbar \omega \approx 11.5$ meV) contributes the most
to the local heating above $V=11.5$ mV, even though the two transverse modes
have slightly smaller frequencies ($\hbar \omega \approx 10.8$ meV) and are
excited at a lower bias. In linear response ($V<10$ mV), two modes
contribute to local heating in the case of the single-molecule junction (see
Fig.~\ref{modes}). 
One, with energy $\hbar \omega \approx 6$ meV, corresponds to the central
benzene ring moving up and down with respect to its equilibrium position
(Fig.~\ref{modes}(a)). The second, with energy $\hbar \omega \approx 9$ meV,
corresponds to the central ring moving sideways with respect to its
equilibrium position (Fig.~\ref{modes}(b)). The first mode contributes to
local heating almost three orders of magnitude more than the second one as
determined by the transformation matrix $A_{i\mu ,j\nu }$. At a bias of
about 0.5V, all vibrational modes of the molecular junction are excited. The
mode that contributes the most to the heating corresponds to an in-plane
``breathing'' of the central carbon ring (not shown).~\cite{mioprlforce}

\begin{figure}[tbp]
\includegraphics[width=.48\textwidth]{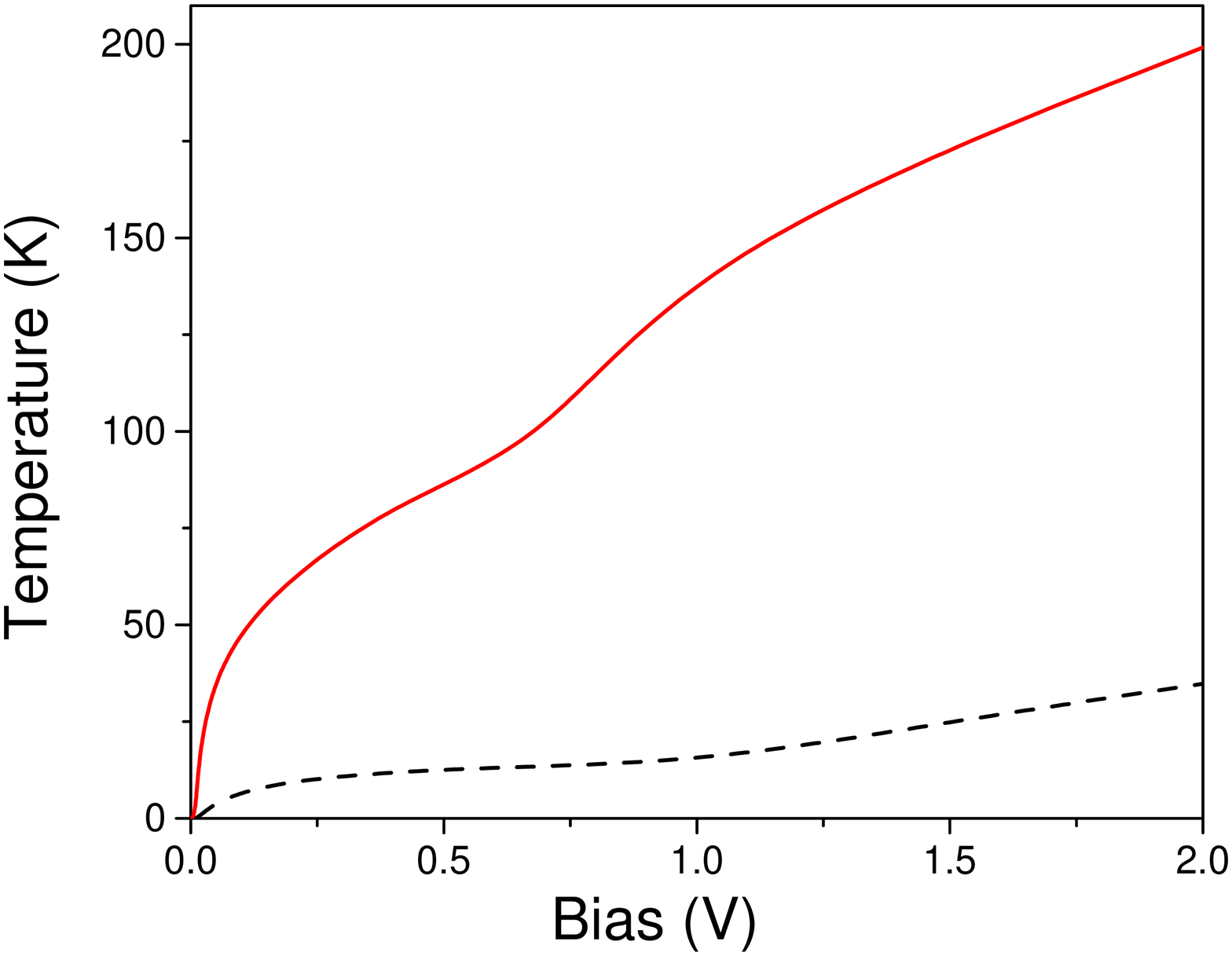}
\caption{Local temperature as a function of bias for a benzene-dithiolate
molecular junction (dashed line) and a gold point contact (solid line) due
to the equilibrium between local heating and heat dissipation into the
electrodes at $T_{e}=0$ K.}
\label{loctemp}
\end{figure}

So far we have assumed the junction is thermally isolated. However, the
majority of the heat generated in the junction is actually transferred to
the electrodes. We estimate the thermal conductance following the approach
of Patton and Geller.~\cite{geller} We assume the junction to be a weak
thermal link with a given stiffness $K$, which we evaluate from first
principles.~\cite{prec1} We then estimate the thermal current into the
electrodes via elastic phonon scattering as~\cite{geller} 
\begin{equation}
I_{th}=\frac{4\pi K^{2}}{\hbar }\int d\varepsilon \varepsilon N_{L}\left(
\varepsilon \right) N_{R}\left( \varepsilon \right) \left[ n_{L}(\varepsilon
)-n_{R}(\varepsilon )\right] ,  \label{dissip}
\end{equation}%
where $n_{L(R)}$ is the Bose-Einstein distribution function, $N_{L(R)}\left(
\varepsilon \right) $ is the spectral density of states at the left (right)
electrode surface, and $K=\pi d^{2}Y/(4l)$. $Y$ is the Young's modulus of
the junction, and $d$ ($l$) is its diameter (length).~\cite{geller} We note
that the temperature profile along a nanojunction is nearly constant and
almost equal to the average temperature of the thermal reservoirs.~\cite%
{talkner,dhar,ciraci,miyashita} The thermal current from the junction with
temperature $T_{w}$ dissipated to the left electrode with temperature $T_{e}$
is thus equivalent to the thermal current of a weak thermal link between
reservoirs with temperatures $T_{e}$ and $2T_{w}-T_{e}$. Analogously for the
thermal current into the right electrode. The effective local temperature,
resulting from the equilibrium between local heating and heat dissipation
into the electrodes, is plotted in Fig.~\ref{loctemp} at $T_{e}=0$~K. Values
of the local temperature for the gold contact are in good agreement with
experimental results~\cite{van} and previous theoretical estimates.~\cite%
{todorov1} 
It is evident from Fig.~\ref{loctemp} that the local temperature is
considerably reduced by heat transfer into the electrodes even for
relatively large biases (cf. Fig.~\ref{eqtemp}). It is also evident that at
any given bias the single-molecule junction heats up less than the gold
point contact. This is due to the larger stiffness of the molecule and its
larger resistance to electrical current.

We thank T.N. Todorov for useful discussions. We acknowledge support from
the NSF Grant Nos. DMR-01-02277 and DMR-01-33075, and Carilion Biomedical
Institute. Acknowledgement is also made to the Donors of The Petroleum
Research Fund, administered by the American Chemical Society, for partial
support of this research.

\end{document}